\documentclass[twocolumn,preprintnumbers,amsmath,amssymb]{revtex4}  

\usepackage{graphicx}
\usepackage{dcolumn}
\usepackage{bm}
\usepackage{multirow}
\usepackage{color}
\usepackage[colorlinks=true]{hyperref}

\begin{document}

\title{Charge tuning in [111] grown GaAs droplet quantum dots }

\author{L. Bouet$^1$}
\author{M. Vidal$^1$}
\author{T. Mano$^2$}
\author{N. Ha$^2$}
\author{T. Kuroda$^2$}
\author{M.~V.~Durnev$^3$}
\author{M.~M.~Glazov$^3$}
\author{E.~L.~Ivchenko$^3$}
\author{X. Marie$^1$}
\author{T. Amand$^1$}
\author{K. Sakoda$^2$}
\author{G. Wang$^1$}
\author{B. Urbaszek$^1$}

\affiliation{%
$^1$Universit\'e de Toulouse, INSA-CNRS-UPS, LPCNO, 135 Av. Rangueil, 31077 Toulouse, France}

\affiliation{%
$^2$National Institute for Material Science, Namiki 1-1, Tsukuba 305-0044, Japan}

\affiliation{%
$^3$Ioffe Physical-Technical Institute RAS, 194021 St.-Petersburg, Russia}


\begin{abstract}
We demonstrate charge tuning in strain free GaAs/AlGaAs quantum dots (QDs) grown by droplet epitaxy on a GaAs(111)A substrate.  Application of a bias voltage allows the controlled charging of the QDs from $-3|e|$ to $+2|e|$. The resulting changes in QD emission energy and exciton fine-structure are recorded in micro-photoluminescence experiments at $T=4$~K. We uncover the existence of excited valence and conduction states, in addition to the $s$-shell-like ground state. We record a second series of emission lines about 25~meV above the charged exciton emission coming from excited charged excitons. For these excited interband transitions a negative diamagnetic shift of large amplitude is uncovered in longitudinal magnetic fields.  
\end{abstract}

\maketitle
\textbf{Introduction.}---
Semiconductor quantum dots (QDs) are true quantum emitters due to their nano-metric size in all three spatial dimensions and QD devices are currently developed for quantum optics and single spin memories \cite{Nowak:2014a,Luxmoore:2013a,Nilsson:2013b,Degreve:2012a,Kodriano:2012a}. 
QD devices grown in the strain-driven Stranski-Krastanov (SK) mode are inherently limited to 
certain combinations of dot-barrier-substrate materials. Droplet epitaxy allows a more flexible approach \cite{Koguchi:1991a,Wang:2007a,Kumah:2009a,Graf:2014a} and GaAs based single photon emitters on Si substrates have been demonstrated \cite{Cavigli:2012a}. Another important advantage compared to SK growth is the free choice of substrate orientation: Here growth along the $[111]$ axis allows the fabrication of QDs with high symmetry \cite{Mano:2010a,Stock:2010a,Treu:2012a}, which are very efficient sources of entangled photon pairs \cite{Kuroda:2013a,Juska:2013a}. Droplet epitaxy on InP(111)A substrates has yielded highly symmetric QD emitters for telecommunication \cite{Ha:2014a,Liu:2014a}.\\
\begin{figure}
\includegraphics[width=0.4\textwidth]{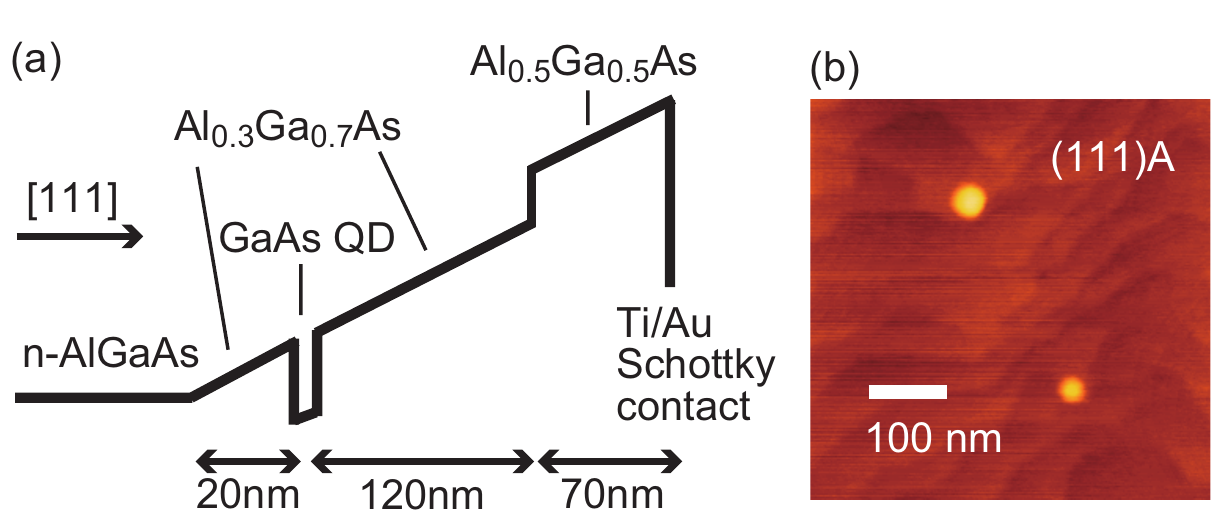}
\caption{\label{fig:fig_scheme}(a) Schematic of the charge-tuning device based on droplet GaAs dots. (b) AFM image of the dots on an AlGaAs(111)A surface.}
\end{figure}
\indent A crucial ingredient for the recent QD device breakthroughs \cite{Gao:2012a,Schaibley:2013a,Gerardot:2008a,Krenner:2006a} is the integration of QDs in field-effect structures (Schottky diodes). There the application of an external electric field allows (i) tuning the QD charge state i.e. the emission polarization and energy \cite{Warburton:2000a,Jovanov:2011a,Ediger:2007a,Ware:2005a,Sanada:2009a}, (ii) strongly suppressing random charge fluctuations, detrimental for device performance such as entanglement fidelity, (iii) operating single QD spin memory devices based on the electrically controlled filling/emptying of the QD with single carriers \cite{Kroutvar:2004a,Maletinsky:2009a}. Controlled charge tuning has been lacking in the promising droplet QD systems so far. \\
\indent In this work we demonstrate the electrically controlled charging of individual, symmetric GaAs/AlGaAs quantum dots grown on a (111)A substrates by droplet epitaxy, Fig.~\ref{fig:fig_scheme}. 
We observe discrete charging steps in photoluminescence (PL) emission from the doubly positively charged exciton X$^{2+}$ up to the triply negatively charged exciton X$^{3-}$, Fig.~\ref{fig:fig1}b~\cite{Karlsson2014}. We extract the electron-hole and electron-electron Coulomb exchange energies from the fine-structure, demonstrating strong carrier confinement, a major advantage compared to GaAs/AlGaAs interface fluctuations dots \cite{Bracker:2005a}. We confirm the existence of 
excited valence and conduction states, in addition to the $s$-shell-like ground state. 
We record a series of excited exciton emission lines about 25~meV above the X$^{2+}$ to X$^{3-}$ series, most prominent when the dot is charged with $\geq3$ electrons. 
For these interband transitions a strongly enhanced, negative diamagnetic shift is uncovered, indicating a wide spatial extension of the wavefunction. 

\begin{figure}
\includegraphics[width=0.46\textwidth]{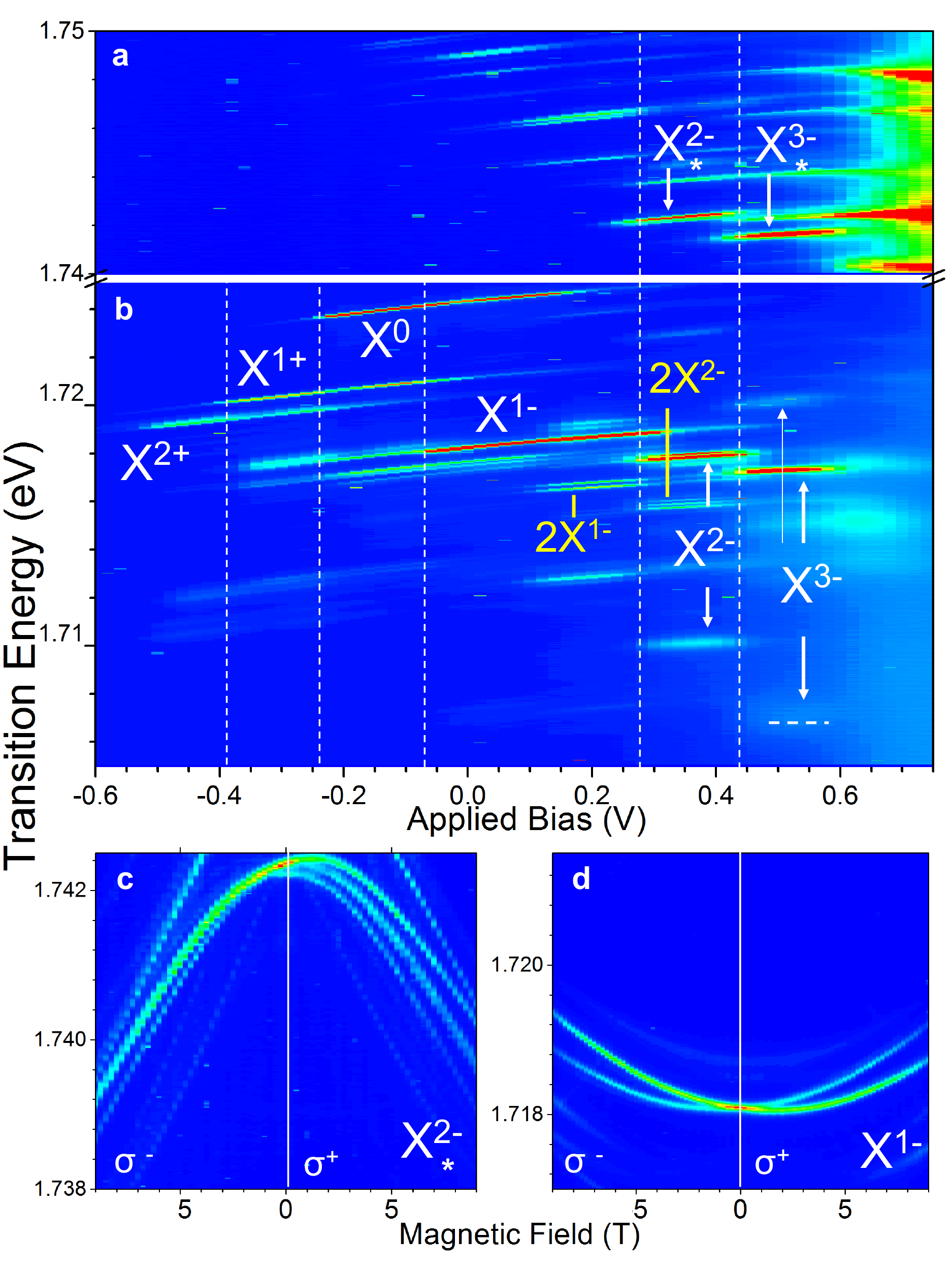}
\caption{\label{fig:fig1} Contour plot of the single dot PL at $T=4$~K as a function of the applied bias voltage.  (a) PL emission involving 
excited states. Blue $<50$ counts, red $>1000$ counts. (b) Main exciton transitions are indicated, where the superscript indicates number and charge of the excess carriers. Discrete charging events are indicated by vertical lines. Blue $<50$ counts, red $>10000$ counts. (c) PL emission in a longitudinal field analyzed in $\sigma^+$ and $\sigma^-$ basis. A large, negative diamagnetic shift of $\gamma(X^{2-}_{*})=-39$~$\mu$ 
eV$\cdot$T$^{-2}$ (d) as (c) but for X$^{1-}$. Conventional, positive diamagnetic shift of $\gamma(X^{1-})=+9$~$\mu$eV$\cdot$T$^{-2}$.}
\end{figure} 

\textbf{Device Fabrication and Experimental Set-up}.--- The sample was grown by droplet epitaxy using a standard molecular beam epitaxy system \cite{Mano:2010a,Sallen:2011a}. 
Starting from the $n^+$-GaAs(111)A substrate, the sample consists of 50-nm $n$-GaAs (Si: $1\times10^{18}$~cm$^{-3}$), 100-nm $n$-Al$_{0.3}$Ga$_{0.7}$As (Si: $1\times10^{18}$~cm$^{-3}$), 20-nm Al$_{0.3}$Ga$_{0.7}$As tunnel barrier, GaAs QDs, 120-nm Al$_{0.3}$Ga$_{0.7}$As, 70-nm Al$_{0.5}$Ga$_{0.5}$As, and 10-nm GaAs cap. The schematic of this structure is illustrated in fig.\ref{fig:fig_scheme}a. For morphology analysis, we deposited an additional sequence of a thin AlGaAs (0.6~nm) followed by the same QD layer without a cap. Figure 1b shows the atomic force microscopy (AFM) image of the surface. It reveals the formation of symmetric dots with a typical height of $\simeq$2-3nm and a radius of $\simeq$15nm. In this model system dots are truly isolated as they are not connected by a 
2D wetting layer
\cite{Mano:2010a,Sallen:2014a}, contrary to SK dots and dots formed at quantum well interface fluctuations \cite{Bracker:2005a}. A semitransparent Ti/Au layer with a nominal thickness of 6 nm serves as a Schottky top gate.\\
\indent A critical factor for this device is the growth of high-quality {\it n}-type barriers. Compared with well-established Si doping in GaAs(100), Si doping in GaAs and AlGaAs(111)A is less stable, where Si can act both as a donor and as an acceptor, and its incorporation behavior depends crucially on the growth conditions -- very high V/III ratios and low growth temperatures are required for $n$-doping \cite{sato:1996}. Moreover, the high crystalline quality of Si:AlGaAs is essential to avoid the emergence of strong deep-level emission, which masks the QD emission. Thus, we carefully optimized the growth conditions in advance, and found successful $n$-type doping at an As$_4$ pressure as high as $\sim8\times10^{-5}$~Torr and a substrate temperature as low as $\sim500^{\circ}$C for a growth rate of 0.2~$\mu$m/h. Hall measurements revealed {\it n} type characters for both Si:GaAs and Si:AlGaAs at 77 K.\\
\indent Single dot PL at 4K is recorded with a home build confocal microscope with a detection spot diameter of
$\simeq 1$~$\mu$m \cite{Durnev:2013a,Sallen:2014a}. The detected PL signal is dispersed by a spectrometer and detected by a Si-CCD
camera (spectral precision of 1~$\mu$eV). Optical excitation is achieved by pumping the AlGaAs barrier with a HeNe laser at 1.96~eV. Laser polarization control and PL polarization analysis is performed with Glan-Taylor polarisers and liquid crystal waveplates. Magnetic fields up to 9~T can be applied parallel (Faraday geometry) to the growth axis [111] that is also the angular momentum quantization axis and the light propagation axis.
\begin{figure}
\includegraphics[width=0.48\textwidth]{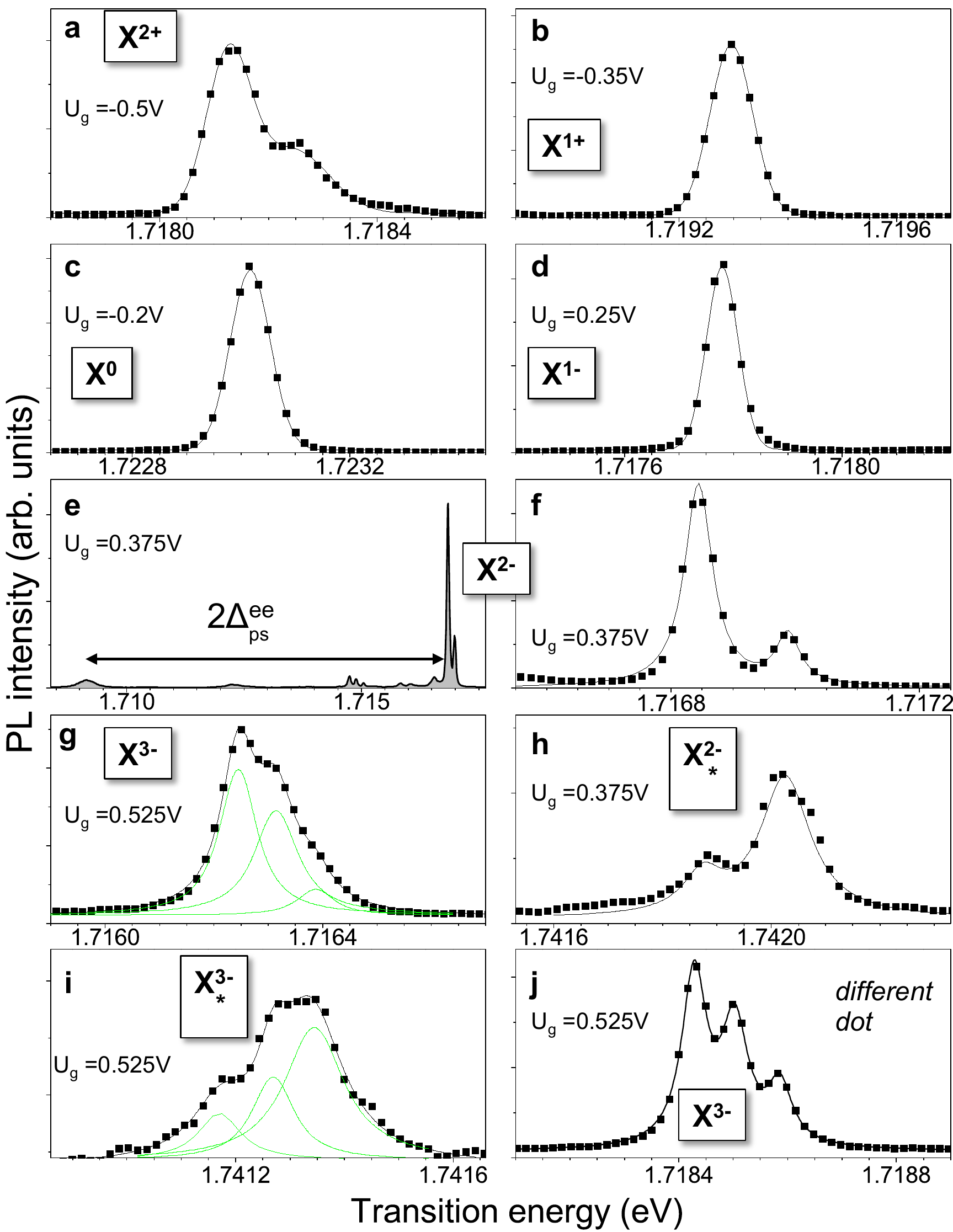}
\caption{\label{fig:fig2} Single dot PL spectra at $T=4$~K for different values of the applied bias $U_g$. Spectra (a) to (i) all from the same QD (a) X$^{2+}$ emission, (b) X$^{1+}$emission, (c) neutral exciton X$^0$ emission, (d) X$^{1-}$emission, (e) X$^{2-}$ emission showing a sharp double peak and a broad singlet peak, separated by $2\Delta^{ee}_{ps}$, (f) Zoom on X$^{2-}$ triplet emission (appears as bright doublet), (g) X$^{3-}$emission, (h) X$^{2-}_{*}$ emission, (i) X$^{3-}_{*}$ emission, (j) X$^{3-}$ emission from \textit{a different dot} with clear triplet emission.}. 
\end{figure} 

\textbf{Results and Discussion}.--- 
In Fig.~\ref{fig:fig1} we show the contour plot of the single dot PL as a function of the applied bias $U_g$. The behaviour is typical for the tens of dots analysed in this device. The charge state identification relies on the characteristic fine structure of the exciton complexes. 
The neutral exciton X$^0$ dominates the PL spectrum for $-0.25$~V$\leq U_g<-0.07$~V. 
This bias range is ideal to further optimize the performance of entangled photon sources based on GaAs droplet dots \cite{Kuroda:2013a}, so far operating in the presence of arbitrary charge fluctuations. QDs grown along $[111]$ are highly symmetric ($C_{3v}$ point symmetry group~\cite{Sallen:2011a,Durnev:2013a}) and we are unable to resolve a splitting $\leq1~\mu$eV for the X$^0$ in the linear basis for the dot shown in Fig.~\ref{fig:fig1}. When further increasing the bias an additional electron is added to the dot and the strong X$^{1-}$ PL emission (see Fig.~\ref{fig:fig2}d), which has no dark states, is recorded for a wide gate voltage range $-0.07~\text{V}\leq U_g \leq+0.30~\text{V}$. This bias range is of particular interest for generating entangled photon-electron spin states \cite{Degreve:2012a,Gao:2012a,Schaibley:2013a}. \\
\indent At $U_g=0.3$~V the X$^{1-}$ emission disappears and a bright doublet emission is observed, accompanied by a broad emission about 7~meV at lower energy, Fig.~\ref{fig:fig2}e,f. This is characteristic for the doubly negatively charged exciton X$^{2-}$ \cite{Urbaszek:2003a}:  in addition to the electron spin singlet on the $s$-shell-level an electron is added to the next excited state denoted as $p$-shell state~\cite{states:trig}.
 The X$^{2-}$ emission gives access to key parameters of this novel QD system: (i) we uncover the existence of an excited
 confined state, (ii)  the energy splitting of the bright doublet allows to extract the electron-hole Coulomb exchange energy $\Delta^{eh}_{ps}$, (iii) the separation between bright doublet and the broader singlet emission is a measure of the electron-electron Coulomb exchange energy $\Delta^{ee}_{ps}$, clearly one order of magnitude larger then $\Delta^{eh}_{ps}$. \\
\indent Further information can be extracted for $U_g\geq0.43$~V as an additional electron is added to the dot to form the triply charged exciton X$^{3-}$. We observe a characteristic triplet structure in Fig.~\ref{fig:fig2}g, even clearer in the example from another dot in Fig.~\ref{fig:fig2}j. The presence of this sharp X$^{3-}$ emission is direct evidence for the high, $C_{3v}$, symmetry of the dot: In QDs with shape asymmetry, $C_{2v}$ or lower, the electron $p_x$ and $p_y$ states are separated in energy. If this separation is larger than electron-electron exchange energy $\Delta^{ee}_{ps}$, the narrow X$^{3-}$ emission is replaced by a broad peak \cite{Karrai:2004a}, which is \textit{not} the case in our experiment. 
In addition to the charged excitons, we also observe charged biexcitons 2X$^{1-}$ and 2X$^{2-}$ in Fig.~\ref{fig:fig1}b, which allows us to extract precise values of the electron-hole Coulomb exchange energies \cite{Urbaszek:2003a}.  Using the simple theory of Ref. \cite{Urbaszek:2003a}, we find an average value over five dots, $\Delta^{eh}_{ss}=325~\mu$eV. For GaAs bulk $\Delta^{eh}\sim20~\mu$eV \cite{Ekardt:1979a}, so the enhancement by more than one order of magnitude for $\Delta^{eh}_{ss}$ is due to the strong carrier confinement in our dots. For the electron-electron Coulomb exchange energy we extract $\Delta^{ee}_{ps}=3.5~$meV. This means the energy cost for the spin flip of a conduction electron is $2\Delta^{ee}_{ps}=7~$meV, which is an important parameter for the carrier spin dynamics in this system~\cite{Sallen:2014a,Laurent:2006a}. \\
\indent For negative bias, PL emission is quenched for $U_g<-0.5$~V, the first peak emerging at $U_g=-0.5$~V  is attributed to the doubly positively charged exciton X$^{2+}$, where the excess holes are photogenerated. The PL emission shows a doublet, Fig.~\ref{fig:fig2}a, that has its origin in electron-hole exchange in the initial state. The X$^{2+}$ contains a total of 3 holes, as the $s$-shell can accommodate only 2-holes, the observation of the X$^{2+}$ emission is a direct proof of the existence of an excited confined state for holes in the dot. For $U_g\geq-0.37$~V a strong, single PL line corresponding to the X$^{1+}$ emerges, Fig.~\ref{fig:fig2}b.\\
\indent The PL emission of X$^{2-}$ (X$^{3-}$) is observed when 3 (4) electrons are in the dot and a photo-generated hole. The presence of both $s$-shell and excited electrons results in a rich fine structure, that is commonly probed via the recombination of the electron $s$-state with the hole $s$-state. Here we report an additional series of charging events for this bias range about 25~meV above the charged exciton series of Fig.~\ref{fig:fig1}b, where we label the most intense transitions X$^{2-}_{*}$ and  X$^{3-}_{*}$ in Fig.~\ref{fig:fig1}a. Exactly in the voltage X$^{2-}$ range $0.3~\text{V}\leq U_g \leq+0.43~\text{V}$ we observe the X$^{2-}_{*}$ emission about an order of magnitude weaker than the main X$^{2-}$ PL peak. This X$^{2-}_{*}$ emission, Fig.~\ref{fig:fig2}h, is split into two lines, separated by a similar energy (within our resolution) as the X$^{2-}$ , Fig.~\ref{fig:fig2}f. Also for the X$^{3-}$ bias range we observe a high energy PL emission in Fig.~\ref{fig:fig1}a. Again the X$^{3-}_{*}$ exhibits a fine structure very similar to the main  X$^{3-}$ emission, cf. Figs.~\ref{fig:fig2}g and \ref{fig:fig2}i. To identify possible origins of these states we assume that the number of electrons is fixed by the bias. 
Most likely, the X$^{2-}_*$ and X$^{3-}_*$  emission lines belong to the same complex, X$^{2-}$ and X$^{3-}$, and involve recombination of an electron in a $p$-shell state and a hole in an $s$-shell, such a transition is possible due to mixing of heavy-hole $s$-shell and light-hole $p$-shell states by off-diagonal terms of Luttinger Hamiltonian due to the polar $C_{3v}$ symmetry of our QDs, cf. Ref.~\cite{Durnev:2013a}. Another option is that these lines correspond to metastable states which involve excited hole states.\\
\indent To further probe the nature of the high energy emission, we have performed measurements in a longitudinal magnetic field $B$ up to 9 Tesla. 
The laser excitation polarization is linear to minimize nuclear spin effects \cite{Urbaszek:2013a} and the PL emission is analysed in the circular basis, energetically separating the Zeeman components. For the  X$^{1-}$ complex, we find the typical behaviour for GaAs droplet QDs with trigonal symmetry grown along the $[111]$ axis: We record four emission lines (2 pairs of $\sigma^+$ and 2 $\sigma^-$ polarized) as dark and bright states are mixed due to heavy-hole coupling \cite{Sallen:2011a,Durnev:2013a}. We extract a diamagnetic shift of $\gamma(\mathrm X^{1-})=+9$~$\mu$eV$\cdot$T$^{-2}$, an additional sign of the strong spatial confinement compared to interface fluctuation QDs \cite{Sanada:2009a}. The PL energy of the  X$^{2-}_{*}$ is plotted as a function of magnetic field in Fig.~\ref{fig:fig1}c. We can distinguish four lines emerging in each polarization $\sigma^+$ and $\sigma^-$ as the field increases. This indicates that the emission is associated to electron-hole recombination with a substantial heavy hole component because transitions involving 
$s$-shell light holes would show only two lines per polarization \cite{Durnev:2013a}. Importantly we observe a striking change of both the sign of the diamagnetic shift and its amplitude with $\gamma(\mathrm X^{2-}_{*})=-39$~$\mu$eV$\cdot$T$^{-2}$, comparing Figs.~\ref{fig:fig1}c and d.\\
\indent The drastic decrease of the PL emission energy as a function of the applied longitudinal magnetic field as well as the negative diamagnetic shift documented in Fig.~ \ref{fig:fig1}c are very surprising and will be a subject of further experimental and theoretical studies. We note here that the strong Coulomb interactions, as documented by the high values of the measured exchange energies in our QDs, are expected to play a role \cite{Schulhauser:2002a}. Also the involvement of excited hole states can result in a sign reversal of the diamagnetic shift~\cite{semina}.\\
\indent To conclude, we have demonstrated charge tuning in [111] GaAs dots grown by droplet epitaxy. The high symmetry of the dots is confirmed by the characteristic X$^{3-}$ emission and negligible fine structure due to anisotropic Coulomb exchange. Device applications such as entangled photon sources \cite{Kuroda:2013a}, spin-photon entanglement \cite{Degreve:2012a,Gao:2012a,Schaibley:2013a} and spin memories using droplet dots can be explored in the absence of charge fluctuations, that limit entanglement fidelity and nuclear spin polarization lifetime in the QD \cite{Sallen:2014a,Urbaszek:2013a}. The demonstration of charge tuning will in addition allow to apply in the future novel experimental techniques to droplet dots such as resonant laser scattering \cite{Hogele:2012a}. We record highly unusual excited state emission, that show strong, negative diamagnetic shifts which probe the dot and wavefunction symmetry. Here further theoretical work is needed to establish the exact nature of these interband transitions.\\
\indent We acknowledge funding from ERC Grant No. 306719. MVD, MMG and ELI were partially supported by Russian Science Foundation (project 14-12-01067), MVD acknowledges support from Dynasty Foundation.


\begin{thebibliography}{43}
\expandafter\ifx\csname natexlab\endcsname\relax\def\natexlab#1{#1}\fi
\expandafter\ifx\csname bibnamefont\endcsname\relax
  \def\bibnamefont#1{#1}\fi
\expandafter\ifx\csname bibfnamefont\endcsname\relax
  \def\bibfnamefont#1{#1}\fi
\expandafter\ifx\csname citenamefont\endcsname\relax
  \def\citenamefont#1{#1}\fi
\expandafter\ifx\csname url\endcsname\relax
  \def\url#1{\texttt{#1}}\fi
\expandafter\ifx\csname urlprefix\endcsname\relax\def\urlprefix{URL }\fi
\providecommand{\bibinfo}[2]{#2}
\providecommand{\eprint}[2][]{\url{#2}}

\bibitem[{\citenamefont{Nowak et~al.}(2014)\citenamefont{Nowak, Portalupi,
  Giesz, Gazzano, Savio, Braun, Karrai, Arnold, Lanco, Sagnes
  et~al.}}]{Nowak:2014a}
\bibinfo{author}{\bibfnamefont{A.~K.} \bibnamefont{Nowak}},
  \bibinfo{author}{\bibfnamefont{S.~L.} \bibnamefont{Portalupi}},
  \bibinfo{author}{\bibfnamefont{V.}~\bibnamefont{Giesz}},
  \bibinfo{author}{\bibfnamefont{O.}~\bibnamefont{Gazzano}},
  \bibinfo{author}{\bibfnamefont{C.~D.} \bibnamefont{Savio}},
  \bibinfo{author}{\bibfnamefont{P.-F.} \bibnamefont{Braun}},
  \bibinfo{author}{\bibfnamefont{K.}~\bibnamefont{Karrai}},
  \bibinfo{author}{\bibfnamefont{C.}~\bibnamefont{Arnold}},
  \bibinfo{author}{\bibfnamefont{L.}~\bibnamefont{Lanco}},
  \bibinfo{author}{\bibfnamefont{I.}~\bibnamefont{Sagnes}},
  \bibnamefont{et~al.}, \bibinfo{journal}{Nature Comms.}
  \textbf{\bibinfo{volume}{5}}, \bibinfo{pages}{3240} (\bibinfo{year}{2014}).

\bibitem[{\citenamefont{Luxmoore et~al.}(2013)\citenamefont{Luxmoore, Wasley,
  Ramsay, Thijssen, Oulton, Hugues, Kasture, Achanta, Fox, and
  Skolnick}}]{Luxmoore:2013a}
\bibinfo{author}{\bibfnamefont{I.~J.} \bibnamefont{Luxmoore}},
  \bibinfo{author}{\bibfnamefont{N.~A.} \bibnamefont{Wasley}},
  \bibinfo{author}{\bibfnamefont{A.~J.} \bibnamefont{Ramsay}},
  \bibinfo{author}{\bibfnamefont{A.~C.~T.} \bibnamefont{Thijssen}},
  \bibinfo{author}{\bibfnamefont{R.}~\bibnamefont{Oulton}},
  \bibinfo{author}{\bibfnamefont{M.}~\bibnamefont{Hugues}},
  \bibinfo{author}{\bibfnamefont{S.}~\bibnamefont{Kasture}},
  \bibinfo{author}{\bibfnamefont{V.~G.} \bibnamefont{Achanta}},
  \bibinfo{author}{\bibfnamefont{A.~M.} \bibnamefont{Fox}}, \bibnamefont{and}
  \bibinfo{author}{\bibfnamefont{M.~S.} \bibnamefont{Skolnick}},
  \bibinfo{journal}{Phys. Rev. Lett.} \textbf{\bibinfo{volume}{110}},
  \bibinfo{pages}{037402} (\bibinfo{year}{2013}).

\bibitem[{\citenamefont{Nilsson et~al.}(2013)\citenamefont{Nilsson, Stevenson,
  Chan, Skiba-Szymanska, Lucamarini, Ward, Bennett, Salter, Farrer, Ritchie
  et~al.}}]{Nilsson:2013b}
\bibinfo{author}{\bibfnamefont{J.}~\bibnamefont{Nilsson}},
  \bibinfo{author}{\bibfnamefont{R.~M.} \bibnamefont{Stevenson}},
  \bibinfo{author}{\bibfnamefont{K.~H.~A.} \bibnamefont{Chan}},
  \bibinfo{author}{\bibfnamefont{J.}~\bibnamefont{Skiba-Szymanska}},
  \bibinfo{author}{\bibfnamefont{M.}~\bibnamefont{Lucamarini}},
  \bibinfo{author}{\bibfnamefont{M.~B.} \bibnamefont{Ward}},
  \bibinfo{author}{\bibfnamefont{A.~J.} \bibnamefont{Bennett}},
  \bibinfo{author}{\bibfnamefont{C.~L.} \bibnamefont{Salter}},
  \bibinfo{author}{\bibfnamefont{I.}~\bibnamefont{Farrer}},
  \bibinfo{author}{\bibfnamefont{D.~A.} \bibnamefont{Ritchie}},
  \bibnamefont{et~al.}, \bibinfo{journal}{Nature Photonics}
  \textbf{\bibinfo{volume}{7}}, \bibinfo{pages}{311} (\bibinfo{year}{2013}).

\bibitem[{\citenamefont{Greve et~al.}(2012)}]{Degreve:2012a}
\bibinfo{author}{\bibfnamefont{K.~D.} \bibnamefont{Greve}}
  \bibnamefont{et~al.}, \bibinfo{journal}{Nature}
  \textbf{\bibinfo{volume}{491}}, \bibinfo{pages}{421} (\bibinfo{year}{2012}).

\bibitem[{\citenamefont{Kodriano et~al.}(2012)\citenamefont{Kodriano, Schwartz,
  Poem, Benny, Presman, Truong, Petroff, and Gershoni}}]{Kodriano:2012a}
\bibinfo{author}{\bibfnamefont{Y.}~\bibnamefont{Kodriano}},
  \bibinfo{author}{\bibfnamefont{I.}~\bibnamefont{Schwartz}},
  \bibinfo{author}{\bibfnamefont{E.}~\bibnamefont{Poem}},
  \bibinfo{author}{\bibfnamefont{Y.}~\bibnamefont{Benny}},
  \bibinfo{author}{\bibfnamefont{R.}~\bibnamefont{Presman}},
  \bibinfo{author}{\bibfnamefont{T.~A.} \bibnamefont{Truong}},
  \bibinfo{author}{\bibfnamefont{P.~M.} \bibnamefont{Petroff}},
  \bibnamefont{and} \bibinfo{author}{\bibfnamefont{D.}~\bibnamefont{Gershoni}},
  \bibinfo{journal}{Phys. Rev. B} \textbf{\bibinfo{volume}{85}},
  \bibinfo{pages}{241304} (\bibinfo{year}{2012}).

\bibitem[{\citenamefont{Koguchi et~al.}(1991)\citenamefont{Koguchi, Takahashi,
  and Chikyow}}]{Koguchi:1991a}
\bibinfo{author}{\bibfnamefont{N.}~\bibnamefont{Koguchi}},
  \bibinfo{author}{\bibfnamefont{S.}~\bibnamefont{Takahashi}},
  \bibnamefont{and} \bibinfo{author}{\bibfnamefont{T.}~\bibnamefont{Chikyow}},
  \bibinfo{journal}{J. Cryst. Growth} \textbf{\bibinfo{volume}{111}},
  \bibinfo{pages}{688} (\bibinfo{year}{1991}).

\bibitem[{\citenamefont{Wang et~al.}(2007)\citenamefont{Wang, Liang, Sablon,
  Lee, Mazur, Strom, and Salamo}}]{Wang:2007a}
\bibinfo{author}{\bibfnamefont{Z.}~\bibnamefont{Wang}},
  \bibinfo{author}{\bibfnamefont{B.}~\bibnamefont{Liang}},
  \bibinfo{author}{\bibfnamefont{K.}~\bibnamefont{Sablon}},
  \bibinfo{author}{\bibfnamefont{J.}~\bibnamefont{Lee}},
  \bibinfo{author}{\bibfnamefont{Y.}~\bibnamefont{Mazur}},
  \bibinfo{author}{\bibfnamefont{N.}~\bibnamefont{Strom}}, \bibnamefont{and}
  \bibinfo{author}{\bibfnamefont{G.}~\bibnamefont{Salamo}},
  \bibinfo{journal}{Small} \textbf{\bibinfo{volume}{3}}, \bibinfo{pages}{235}
  (\bibinfo{year}{2007}), ISSN \bibinfo{issn}{1613-6829}.

\bibitem[{\citenamefont{Kumah et~al.}(2009)\citenamefont{Kumah, Shusterman,
  Paltiel, Yacoby, and Clarke}}]{Kumah:2009a}
\bibinfo{author}{\bibfnamefont{D.~P.} \bibnamefont{Kumah}},
  \bibinfo{author}{\bibfnamefont{S.}~\bibnamefont{Shusterman}},
  \bibinfo{author}{\bibfnamefont{Y.}~\bibnamefont{Paltiel}},
  \bibinfo{author}{\bibfnamefont{Y.}~\bibnamefont{Yacoby}}, \bibnamefont{and}
  \bibinfo{author}{\bibfnamefont{R.}~\bibnamefont{Clarke}},
  \bibinfo{journal}{Nature Nanotech.} \textbf{\bibinfo{volume}{4}},
  \bibinfo{pages}{835} (\bibinfo{year}{2009}).

\bibitem[{\citenamefont{Graf et~al.}(2014)\citenamefont{Graf, Sonnenberg,
  Paulava, Schliwa, Heyn, and Hansen}}]{Graf:2014a}
\bibinfo{author}{\bibfnamefont{A.}~\bibnamefont{Graf}},
  \bibinfo{author}{\bibfnamefont{D.}~\bibnamefont{Sonnenberg}},
  \bibinfo{author}{\bibfnamefont{V.}~\bibnamefont{Paulava}},
  \bibinfo{author}{\bibfnamefont{A.}~\bibnamefont{Schliwa}},
  \bibinfo{author}{\bibfnamefont{C.}~\bibnamefont{Heyn}}, \bibnamefont{and}
  \bibinfo{author}{\bibfnamefont{W.}~\bibnamefont{Hansen}},
  \bibinfo{journal}{Phys. Rev. B} \textbf{\bibinfo{volume}{89}},
  \bibinfo{pages}{115314} (\bibinfo{year}{2014}).

\bibitem[{\citenamefont{Cavigli et~al.}(2012)\citenamefont{Cavigli, Bietti,
  Accanto, Minari, Abbarchi, Isella, Frigeri, Vinattieri, Gurioli, and
  Sanguinetti}}]{Cavigli:2012a}
\bibinfo{author}{\bibfnamefont{L.}~\bibnamefont{Cavigli}},
  \bibinfo{author}{\bibfnamefont{S.}~\bibnamefont{Bietti}},
  \bibinfo{author}{\bibfnamefont{N.}~\bibnamefont{Accanto}},
  \bibinfo{author}{\bibfnamefont{S.}~\bibnamefont{Minari}},
  \bibinfo{author}{\bibfnamefont{M.}~\bibnamefont{Abbarchi}},
  \bibinfo{author}{\bibfnamefont{G.}~\bibnamefont{Isella}},
  \bibinfo{author}{\bibfnamefont{C.}~\bibnamefont{Frigeri}},
  \bibinfo{author}{\bibfnamefont{A.}~\bibnamefont{Vinattieri}},
  \bibinfo{author}{\bibfnamefont{M.}~\bibnamefont{Gurioli}}, \bibnamefont{and}
  \bibinfo{author}{\bibfnamefont{S.}~\bibnamefont{Sanguinetti}},
  \bibinfo{journal}{Applied Physics Letters} \textbf{\bibinfo{volume}{100}},
  \bibinfo{eid}{231112} (\bibinfo{year}{2012}).

\bibitem[{\citenamefont{Mano et~al.}(2010)\citenamefont{Mano, Abbarchi, Kuroda,
  McSkimming, Ohtake, Mitsuishi, and Sakoda}}]{Mano:2010a}
\bibinfo{author}{\bibfnamefont{T.}~\bibnamefont{Mano}},
  \bibinfo{author}{\bibfnamefont{M.}~\bibnamefont{Abbarchi}},
  \bibinfo{author}{\bibfnamefont{T.}~\bibnamefont{Kuroda}},
  \bibinfo{author}{\bibfnamefont{B.}~\bibnamefont{McSkimming}},
  \bibinfo{author}{\bibfnamefont{A.}~\bibnamefont{Ohtake}},
  \bibinfo{author}{\bibfnamefont{K.}~\bibnamefont{Mitsuishi}},
  \bibnamefont{and} \bibinfo{author}{\bibfnamefont{K.}~\bibnamefont{Sakoda}},
  \bibinfo{journal}{Applied Physics Express} \textbf{\bibinfo{volume}{3}},
  \bibinfo{pages}{065203} (\bibinfo{year}{2010}).

\bibitem[{\citenamefont{Stock et~al.}(2010)}]{Stock:2010a}
\bibinfo{author}{\bibfnamefont{E.}~\bibnamefont{Stock}} \bibnamefont{et~al.},
  \bibinfo{journal}{Appl. Phys. Lett.} \textbf{\bibinfo{volume}{96}},
  \bibinfo{eid}{093112} (\bibinfo{year}{2010}).

\bibitem[{\citenamefont{Treu et~al.}(2012)\citenamefont{Treu, Schneider,
  Huggenberger, Braun, Reitzenstein, Höfling, and Kamp}}]{Treu:2012a}
\bibinfo{author}{\bibfnamefont{J.}~\bibnamefont{Treu}},
  \bibinfo{author}{\bibfnamefont{C.}~\bibnamefont{Schneider}},
  \bibinfo{author}{\bibfnamefont{A.}~\bibnamefont{Huggenberger}},
  \bibinfo{author}{\bibfnamefont{T.}~\bibnamefont{Braun}},
  \bibinfo{author}{\bibfnamefont{S.}~\bibnamefont{Reitzenstein}},
  \bibinfo{author}{\bibfnamefont{S.}~\bibnamefont{Höfling}}, \bibnamefont{and}
  \bibinfo{author}{\bibfnamefont{M.}~\bibnamefont{Kamp}},
  \bibinfo{journal}{Applied Physics Letters} \textbf{\bibinfo{volume}{101}},
  \bibinfo{eid}{022102} (\bibinfo{year}{2012}).

\bibitem[{\citenamefont{Kuroda et~al.}(2013)\citenamefont{Kuroda, Mano, Ha,
  Nakajima, Kumano, Urbaszek, Jo, Abbarchi, Sakuma, Sakoda
  et~al.}}]{Kuroda:2013a}
\bibinfo{author}{\bibfnamefont{T.}~\bibnamefont{Kuroda}},
  \bibinfo{author}{\bibfnamefont{T.}~\bibnamefont{Mano}},
  \bibinfo{author}{\bibfnamefont{N.}~\bibnamefont{Ha}},
  \bibinfo{author}{\bibfnamefont{H.}~\bibnamefont{Nakajima}},
  \bibinfo{author}{\bibfnamefont{H.}~\bibnamefont{Kumano}},
  \bibinfo{author}{\bibfnamefont{B.}~\bibnamefont{Urbaszek}},
  \bibinfo{author}{\bibfnamefont{M.}~\bibnamefont{Jo}},
  \bibinfo{author}{\bibfnamefont{M.}~\bibnamefont{Abbarchi}},
  \bibinfo{author}{\bibfnamefont{Y.}~\bibnamefont{Sakuma}},
  \bibinfo{author}{\bibfnamefont{K.}~\bibnamefont{Sakoda}},
  \bibnamefont{et~al.}, \bibinfo{journal}{Phys. Rev. B}
  \textbf{\bibinfo{volume}{88}}, \bibinfo{pages}{041306}
  (\bibinfo{year}{2013}).

\bibitem[{\citenamefont{Juska et~al.}(2013)\citenamefont{Juska, Dimastrodonato,
  Mereni, Gocalinska, and Pelucchi}}]{Juska:2013a}
\bibinfo{author}{\bibfnamefont{G.}~\bibnamefont{Juska}},
  \bibinfo{author}{\bibfnamefont{V.}~\bibnamefont{Dimastrodonato}},
  \bibinfo{author}{\bibfnamefont{L.~O.} \bibnamefont{Mereni}},
  \bibinfo{author}{\bibfnamefont{A.}~\bibnamefont{Gocalinska}},
  \bibnamefont{and} \bibinfo{author}{\bibfnamefont{E.}~\bibnamefont{Pelucchi}},
  \bibinfo{journal}{Nature Photon.} \textbf{\bibinfo{volume}{7}},
  \bibinfo{pages}{527} (\bibinfo{year}{2013}).

\bibitem[{\citenamefont{Ha et~al.}(2014)\citenamefont{Ha, Liu, Mano, Kuroda,
  Mitsuishi, Castellano, Sanguinetti, Noda, Sakuma, and Sakoda}}]{Ha:2014a}
\bibinfo{author}{\bibfnamefont{N.}~\bibnamefont{Ha}},
  \bibinfo{author}{\bibfnamefont{X.}~\bibnamefont{Liu}},
  \bibinfo{author}{\bibfnamefont{T.}~\bibnamefont{Mano}},
  \bibinfo{author}{\bibfnamefont{T.}~\bibnamefont{Kuroda}},
  \bibinfo{author}{\bibfnamefont{K.}~\bibnamefont{Mitsuishi}},
  \bibinfo{author}{\bibfnamefont{A.}~\bibnamefont{Castellano}},
  \bibinfo{author}{\bibfnamefont{S.}~\bibnamefont{Sanguinetti}},
  \bibinfo{author}{\bibfnamefont{T.}~\bibnamefont{Noda}},
  \bibinfo{author}{\bibfnamefont{Y.}~\bibnamefont{Sakuma}}, \bibnamefont{and}
  \bibinfo{author}{\bibfnamefont{K.}~\bibnamefont{Sakoda}},
  \bibinfo{journal}{Applied Physics Letters} \textbf{\bibinfo{volume}{104}},
  \bibinfo{eid}{143106} (\bibinfo{year}{2014}).

\bibitem[{\citenamefont{{Liu} et~al.}(2014)\citenamefont{{Liu}, {Ha},
  {Nakajima}, {Mano}, {Kuroda}, {Urbaszek}, {Kumano}, {Suemune}, {Sakuma}, and
  {Sakoda}}}]{Liu:2014a}
\bibinfo{author}{\bibfnamefont{X.}~\bibnamefont{{Liu}}},
  \bibinfo{author}{\bibfnamefont{N.}~\bibnamefont{{Ha}}},
  \bibinfo{author}{\bibfnamefont{H.}~\bibnamefont{{Nakajima}}},
  \bibinfo{author}{\bibfnamefont{T.}~\bibnamefont{{Mano}}},
  \bibinfo{author}{\bibfnamefont{T.}~\bibnamefont{{Kuroda}}},
  \bibinfo{author}{\bibfnamefont{B.}~\bibnamefont{{Urbaszek}}},
  \bibinfo{author}{\bibfnamefont{H.}~\bibnamefont{{Kumano}}},
  \bibinfo{author}{\bibfnamefont{I.}~\bibnamefont{{Suemune}}},
  \bibinfo{author}{\bibfnamefont{Y.}~\bibnamefont{{Sakuma}}}, \bibnamefont{and}
  \bibinfo{author}{\bibfnamefont{K.}~\bibnamefont{{Sakoda}}},
  \bibinfo{journal}{ArXiv e-prints}  (\bibinfo{year}{2014}),
  \eprint{1406.4576}.

\bibitem[{\citenamefont{Gao et~al.}(2012)}]{Gao:2012a}
\bibinfo{author}{\bibfnamefont{W.~B.} \bibnamefont{Gao}} \bibnamefont{et~al.},
  \bibinfo{journal}{Nature} \textbf{\bibinfo{volume}{491}},
  \bibinfo{pages}{426} (\bibinfo{year}{2012}).

\bibitem[{\citenamefont{Schaibley et~al.}(2013)\citenamefont{Schaibley, Burgers
  et~al.}}]{Schaibley:2013a}
\bibinfo{author}{\bibfnamefont{J.~R.} \bibnamefont{Schaibley}},
  \bibinfo{author}{\bibfnamefont{A.~P.} \bibnamefont{Burgers}},
  \bibnamefont{et~al.}, \bibinfo{journal}{Phys. Rev. Lett.}
  \textbf{\bibinfo{volume}{110}}, \bibinfo{pages}{167401}
  (\bibinfo{year}{2013}).

\bibitem[{\citenamefont{Gerardot et~al.}(2006)\citenamefont{Gerardot, Brunner,
  Dalgarno, Ohberg, Seidl, Kroner, Karrai, G.~Stoltz, Petroff, and
  Warburton}}]{Gerardot:2008a}
\bibinfo{author}{\bibfnamefont{B.~D.} \bibnamefont{Gerardot}},
  \bibinfo{author}{\bibfnamefont{D.}~\bibnamefont{Brunner}},
  \bibinfo{author}{\bibfnamefont{P.~A.} \bibnamefont{Dalgarno}},
  \bibinfo{author}{\bibfnamefont{P.}~\bibnamefont{Ohberg}},
  \bibinfo{author}{\bibfnamefont{S.}~\bibnamefont{Seidl}},
  \bibinfo{author}{\bibfnamefont{M.}~\bibnamefont{Kroner}},
  \bibinfo{author}{\bibfnamefont{K.}~\bibnamefont{Karrai}},
  \bibinfo{author}{\bibfnamefont{N.}~\bibnamefont{G.~Stoltz}},
  \bibinfo{author}{\bibfnamefont{P.~M.} \bibnamefont{Petroff}},
  \bibnamefont{and} \bibinfo{author}{\bibfnamefont{R.~J.}
  \bibnamefont{Warburton}}, \bibinfo{journal}{Nature}
  \textbf{\bibinfo{volume}{451}}, \bibinfo{pages}{441} (\bibinfo{year}{2006}).

\bibitem[{\citenamefont{Krenner et~al.}(2006)\citenamefont{Krenner, Clark,
  Nakaoka, Bichler, Scheurer, Abstreiter, and Finley}}]{Krenner:2006a}
\bibinfo{author}{\bibfnamefont{H.~J.} \bibnamefont{Krenner}},
  \bibinfo{author}{\bibfnamefont{E.~C.} \bibnamefont{Clark}},
  \bibinfo{author}{\bibfnamefont{T.}~\bibnamefont{Nakaoka}},
  \bibinfo{author}{\bibfnamefont{M.}~\bibnamefont{Bichler}},
  \bibinfo{author}{\bibfnamefont{C.}~\bibnamefont{Scheurer}},
  \bibinfo{author}{\bibfnamefont{G.}~\bibnamefont{Abstreiter}},
  \bibnamefont{and} \bibinfo{author}{\bibfnamefont{J.~J.}
  \bibnamefont{Finley}}, \bibinfo{journal}{Phys. Rev. Lett.}
  \textbf{\bibinfo{volume}{97}}, \bibinfo{pages}{076403}
  (\bibinfo{year}{2006}).

\bibitem[{\citenamefont{Warburton et~al.}(2000)\citenamefont{Warburton,
  Schaflein, Haft, Bickel, Lorke, Karrai, Garcia, Schoenfeld, and
  Petroff}}]{Warburton:2000a}
\bibinfo{author}{\bibfnamefont{R.~J.} \bibnamefont{Warburton}},
  \bibinfo{author}{\bibfnamefont{C.}~\bibnamefont{Schaflein}},
  \bibinfo{author}{\bibfnamefont{D.}~\bibnamefont{Haft}},
  \bibinfo{author}{\bibfnamefont{F.}~\bibnamefont{Bickel}},
  \bibinfo{author}{\bibfnamefont{A.}~\bibnamefont{Lorke}},
  \bibinfo{author}{\bibfnamefont{K.}~\bibnamefont{Karrai}},
  \bibinfo{author}{\bibfnamefont{J.~M.} \bibnamefont{Garcia}},
  \bibinfo{author}{\bibfnamefont{W.}~\bibnamefont{Schoenfeld}},
  \bibnamefont{and} \bibinfo{author}{\bibfnamefont{P.~M.}
  \bibnamefont{Petroff}}, \bibinfo{journal}{Nature}
  \textbf{\bibinfo{volume}{405}}, \bibinfo{pages}{926} (\bibinfo{year}{2000}).

\bibitem[{\citenamefont{Jovanov et~al.}(2011)\citenamefont{Jovanov, Kapfinger,
  Bichler, Abstreiter, and Finley}}]{Jovanov:2011a}
\bibinfo{author}{\bibfnamefont{V.}~\bibnamefont{Jovanov}},
  \bibinfo{author}{\bibfnamefont{S.}~\bibnamefont{Kapfinger}},
  \bibinfo{author}{\bibfnamefont{M.}~\bibnamefont{Bichler}},
  \bibinfo{author}{\bibfnamefont{G.}~\bibnamefont{Abstreiter}},
  \bibnamefont{and} \bibinfo{author}{\bibfnamefont{J.~J.}
  \bibnamefont{Finley}}, \bibinfo{journal}{Phys. Rev. B}
  \textbf{\bibinfo{volume}{84}}, \bibinfo{pages}{235321}
  (\bibinfo{year}{2011}).

\bibitem[{\citenamefont{Ediger et~al.}(2007)\citenamefont{Ediger, Bester,
  Badolato, Petroff, Karrai, Zunger, and Warburton}}]{Ediger:2007a}
\bibinfo{author}{\bibfnamefont{M.}~\bibnamefont{Ediger}},
  \bibinfo{author}{\bibfnamefont{G.}~\bibnamefont{Bester}},
  \bibinfo{author}{\bibfnamefont{A.}~\bibnamefont{Badolato}},
  \bibinfo{author}{\bibfnamefont{P.~M.} \bibnamefont{Petroff}},
  \bibinfo{author}{\bibfnamefont{K.}~\bibnamefont{Karrai}},
  \bibinfo{author}{\bibfnamefont{A.}~\bibnamefont{Zunger}}, \bibnamefont{and}
  \bibinfo{author}{\bibfnamefont{R.~J.} \bibnamefont{Warburton}},
  \bibinfo{journal}{Nature Physics} \textbf{\bibinfo{volume}{3}},
  \bibinfo{pages}{774} (\bibinfo{year}{2007}).

\bibitem[{\citenamefont{Ware et~al.}(2005)\citenamefont{Ware, Stinaff, Gammon,
  Doty, Bracker, Gershoni, Korenev, Badescu, Lyanda-Geller, and
  Reinecke}}]{Ware:2005a}
\bibinfo{author}{\bibfnamefont{M.~E.} \bibnamefont{Ware}},
  \bibinfo{author}{\bibfnamefont{E.~A.} \bibnamefont{Stinaff}},
  \bibinfo{author}{\bibfnamefont{D.}~\bibnamefont{Gammon}},
  \bibinfo{author}{\bibfnamefont{M.~F.} \bibnamefont{Doty}},
  \bibinfo{author}{\bibfnamefont{A.~S.} \bibnamefont{Bracker}},
  \bibinfo{author}{\bibfnamefont{D.}~\bibnamefont{Gershoni}},
  \bibinfo{author}{\bibfnamefont{V.~L.} \bibnamefont{Korenev}},
  \bibinfo{author}{\bibfnamefont{S.~C.} \bibnamefont{Badescu}},
  \bibinfo{author}{\bibfnamefont{Y.}~\bibnamefont{Lyanda-Geller}},
  \bibnamefont{and} \bibinfo{author}{\bibfnamefont{T.~L.}
  \bibnamefont{Reinecke}}, \bibinfo{journal}{Phys. Rev. Lett.}
  \textbf{\bibinfo{volume}{95}}, \bibinfo{pages}{177403}
  (\bibinfo{year}{2005}).

\bibitem[{\citenamefont{Sanada et~al.}(2009)\citenamefont{Sanada, Sogawa,
  Gotoh, Tokura, Yamaguchi, Nakano, and Kamada}}]{Sanada:2009a}
\bibinfo{author}{\bibfnamefont{H.}~\bibnamefont{Sanada}},
  \bibinfo{author}{\bibfnamefont{T.}~\bibnamefont{Sogawa}},
  \bibinfo{author}{\bibfnamefont{H.}~\bibnamefont{Gotoh}},
  \bibinfo{author}{\bibfnamefont{Y.}~\bibnamefont{Tokura}},
  \bibinfo{author}{\bibfnamefont{H.}~\bibnamefont{Yamaguchi}},
  \bibinfo{author}{\bibfnamefont{H.}~\bibnamefont{Nakano}}, \bibnamefont{and}
  \bibinfo{author}{\bibfnamefont{H.}~\bibnamefont{Kamada}},
  \bibinfo{journal}{Phys. Rev. B} \textbf{\bibinfo{volume}{79}},
  \bibinfo{pages}{121303} (\bibinfo{year}{2009}).

\bibitem[{\citenamefont{Kroutvar et~al.}(2004)\citenamefont{Kroutvar, Ducommun,
  Heiss, Bichler, Schuh, Abstreiter, and Finley}}]{Kroutvar:2004a}
\bibinfo{author}{\bibfnamefont{M.}~\bibnamefont{Kroutvar}},
  \bibinfo{author}{\bibfnamefont{Y.}~\bibnamefont{Ducommun}},
  \bibinfo{author}{\bibfnamefont{D.}~\bibnamefont{Heiss}},
  \bibinfo{author}{\bibfnamefont{M.}~\bibnamefont{Bichler}},
  \bibinfo{author}{\bibfnamefont{D.}~\bibnamefont{Schuh}},
  \bibinfo{author}{\bibfnamefont{G.}~\bibnamefont{Abstreiter}},
  \bibnamefont{and} \bibinfo{author}{\bibfnamefont{J.~J.}
  \bibnamefont{Finley}}, \bibinfo{journal}{Nature}
  \textbf{\bibinfo{volume}{432}}, \bibinfo{pages}{81} (\bibinfo{year}{2004}).

\bibitem[{\citenamefont{Maletinsky et~al.}(2009)\citenamefont{Maletinsky,
  Kroner, and Imamoglu}}]{Maletinsky:2009a}
\bibinfo{author}{\bibfnamefont{P.}~\bibnamefont{Maletinsky}},
  \bibinfo{author}{\bibfnamefont{M.}~\bibnamefont{Kroner}}, \bibnamefont{and}
  \bibinfo{author}{\bibfnamefont{A.}~\bibnamefont{Imamoglu}},
  \bibinfo{journal}{Nature Phys.} \textbf{\bibinfo{volume}{5}},
  \bibinfo{pages}{407} (\bibinfo{year}{2009}).

\bibitem[{Kar()}]{Karlsson2014}
\bibinfo{note}{Many-particle states in ungated GaAs/AlGaAs QDs grown along the
  [111]B axis were also recently reported by K.F. Karlsson et al.,
  arXiv:1407.4042.}

\bibitem[{\citenamefont{Bracker et~al.}(2005)\citenamefont{Bracker, Stinaff,
  Gammon, Ware, Tischler, Shabaev, Efros, Park, Gershoni, Korenev
  et~al.}}]{Bracker:2005a}
\bibinfo{author}{\bibfnamefont{A.~S.} \bibnamefont{Bracker}},
  \bibinfo{author}{\bibfnamefont{E.~A.} \bibnamefont{Stinaff}},
  \bibinfo{author}{\bibfnamefont{D.}~\bibnamefont{Gammon}},
  \bibinfo{author}{\bibfnamefont{M.~E.} \bibnamefont{Ware}},
  \bibinfo{author}{\bibfnamefont{J.~G.} \bibnamefont{Tischler}},
  \bibinfo{author}{\bibfnamefont{A.}~\bibnamefont{Shabaev}},
  \bibinfo{author}{\bibfnamefont{A.~L.} \bibnamefont{Efros}},
  \bibinfo{author}{\bibfnamefont{D.}~\bibnamefont{Park}},
  \bibinfo{author}{\bibfnamefont{D.}~\bibnamefont{Gershoni}},
  \bibinfo{author}{\bibfnamefont{V.~L.} \bibnamefont{Korenev}},
  \bibnamefont{et~al.}, \bibinfo{journal}{Phys. Rev. Lett.}
  \textbf{\bibinfo{volume}{94}}, \bibinfo{pages}{047402}
  (\bibinfo{year}{2005}).

\bibitem[{\citenamefont{Sallen et~al.}(2011)\citenamefont{Sallen, Urbaszek,
  Glazov, Ivchenko, Kuroda, Mano, Kunz, Abbarchi, Sakoda, Lagarde
  et~al.}}]{Sallen:2011a}
\bibinfo{author}{\bibfnamefont{G.}~\bibnamefont{Sallen}},
  \bibinfo{author}{\bibfnamefont{B.}~\bibnamefont{Urbaszek}},
  \bibinfo{author}{\bibfnamefont{M.~M.} \bibnamefont{Glazov}},
  \bibinfo{author}{\bibfnamefont{E.~L.} \bibnamefont{Ivchenko}},
  \bibinfo{author}{\bibfnamefont{T.}~\bibnamefont{Kuroda}},
  \bibinfo{author}{\bibfnamefont{T.}~\bibnamefont{Mano}},
  \bibinfo{author}{\bibfnamefont{S.}~\bibnamefont{Kunz}},
  \bibinfo{author}{\bibfnamefont{M.}~\bibnamefont{Abbarchi}},
  \bibinfo{author}{\bibfnamefont{K.}~\bibnamefont{Sakoda}},
  \bibinfo{author}{\bibfnamefont{D.}~\bibnamefont{Lagarde}},
  \bibnamefont{et~al.}, \bibinfo{journal}{Phys. Rev. Lett.}
  \textbf{\bibinfo{volume}{107}}, \bibinfo{pages}{166604}
  (\bibinfo{year}{2011}).

\bibitem[{\citenamefont{Sallen et~al.}(2014)\citenamefont{Sallen, Kunz, Amand,
  Bouet, Kuroda, Mano, Paget, Krebs, Marie, Sakoda et~al.}}]{Sallen:2014a}
\bibinfo{author}{\bibfnamefont{G.}~\bibnamefont{Sallen}},
  \bibinfo{author}{\bibfnamefont{S.}~\bibnamefont{Kunz}},
  \bibinfo{author}{\bibfnamefont{T.}~\bibnamefont{Amand}},
  \bibinfo{author}{\bibfnamefont{L.}~\bibnamefont{Bouet}},
  \bibinfo{author}{\bibfnamefont{T.}~\bibnamefont{Kuroda}},
  \bibinfo{author}{\bibfnamefont{T.}~\bibnamefont{Mano}},
  \bibinfo{author}{\bibfnamefont{D.}~\bibnamefont{Paget}},
  \bibinfo{author}{\bibfnamefont{O.}~\bibnamefont{Krebs}},
  \bibinfo{author}{\bibfnamefont{X.}~\bibnamefont{Marie}},
  \bibinfo{author}{\bibfnamefont{K.}~\bibnamefont{Sakoda}},
  \bibnamefont{et~al.}, \bibinfo{journal}{Nature Comms.}
  \textbf{\bibinfo{volume}{5}}, \bibinfo{pages}{3268} (\bibinfo{year}{2014}).

\bibitem[{\citenamefont{Sato et~al.}(1996)\citenamefont{Sato, Fahy, Ashwin, and
  Joyce}}]{sato:1996}
\bibinfo{author}{\bibfnamefont{K.}~\bibnamefont{Sato}},
  \bibinfo{author}{\bibfnamefont{M.~R.} \bibnamefont{Fahy}},
  \bibinfo{author}{\bibfnamefont{M.~J.} \bibnamefont{Ashwin}},
  \bibnamefont{and} \bibinfo{author}{\bibfnamefont{B.~A.} \bibnamefont{Joyce}},
  \bibinfo{journal}{Journal of Crystal Growth} \textbf{\bibinfo{volume}{165}},
  \bibinfo{pages}{345 } (\bibinfo{year}{1996}), ISSN \bibinfo{issn}{0022-0248}.

\bibitem[{\citenamefont{Durnev et~al.}(2013)\citenamefont{Durnev, Glazov,
  Ivchenko, Jo, Mano, Kuroda, Sakoda, Kunz, Sallen, Bouet
  et~al.}}]{Durnev:2013a}
\bibinfo{author}{\bibfnamefont{M.~V.} \bibnamefont{Durnev}},
  \bibinfo{author}{\bibfnamefont{M.~M.} \bibnamefont{Glazov}},
  \bibinfo{author}{\bibfnamefont{E.~L.} \bibnamefont{Ivchenko}},
  \bibinfo{author}{\bibfnamefont{M.}~\bibnamefont{Jo}},
  \bibinfo{author}{\bibfnamefont{T.}~\bibnamefont{Mano}},
  \bibinfo{author}{\bibfnamefont{T.}~\bibnamefont{Kuroda}},
  \bibinfo{author}{\bibfnamefont{K.}~\bibnamefont{Sakoda}},
  \bibinfo{author}{\bibfnamefont{S.}~\bibnamefont{Kunz}},
  \bibinfo{author}{\bibfnamefont{G.}~\bibnamefont{Sallen}},
  \bibinfo{author}{\bibfnamefont{L.}~\bibnamefont{Bouet}},
  \bibnamefont{et~al.}, \bibinfo{journal}{Phys. Rev. B}
  \textbf{\bibinfo{volume}{87}}, \bibinfo{pages}{085315}
  (\bibinfo{year}{2013}).

\bibitem[{\citenamefont{Urbaszek et~al.}(2003)\citenamefont{Urbaszek,
  Warburton, Karrai, Gerardot, Petroff, and Garcia}}]{Urbaszek:2003a}
\bibinfo{author}{\bibfnamefont{B.}~\bibnamefont{Urbaszek}},
  \bibinfo{author}{\bibfnamefont{R.~J.} \bibnamefont{Warburton}},
  \bibinfo{author}{\bibfnamefont{K.}~\bibnamefont{Karrai}},
  \bibinfo{author}{\bibfnamefont{B.~D.} \bibnamefont{Gerardot}},
  \bibinfo{author}{\bibfnamefont{P.~M.} \bibnamefont{Petroff}},
  \bibnamefont{and} \bibinfo{author}{\bibfnamefont{J.~M.}
  \bibnamefont{Garcia}}, \bibinfo{journal}{Phys. Rev. Lett.}
  \textbf{\bibinfo{volume}{90}}, \bibinfo{pages}{247403}
  (\bibinfo{year}{2003}).

\bibitem[{sta()}]{states:trig}
\bibinfo{note}{In the $C_{3v}$ point symmetry group the in-plane envelope
  functions transform either according to the non-degenerate representation
  $\Gamma_1$ ($s$-shell) or to the two-dimensional representation $\Gamma_4$
  ($p$-shell, by analogy with axially symmetric dots).}

\bibitem[{\citenamefont{Karrai et~al.}(2004)}]{Karrai:2004a}
\bibinfo{author}{\bibfnamefont{K.}~\bibnamefont{Karrai}} \bibnamefont{et~al.},
  \bibinfo{journal}{Nature} \textbf{\bibinfo{volume}{427}},
  \bibinfo{pages}{135} (\bibinfo{year}{2004}).

\bibitem[{\citenamefont{Ekardt et~al.}(1979)\citenamefont{Ekardt, L\"osch, and
  Bimberg}}]{Ekardt:1979a}
\bibinfo{author}{\bibfnamefont{W.}~\bibnamefont{Ekardt}},
  \bibinfo{author}{\bibfnamefont{K.}~\bibnamefont{L\"osch}}, \bibnamefont{and}
  \bibinfo{author}{\bibfnamefont{D.}~\bibnamefont{Bimberg}},
  \bibinfo{journal}{Phys. Rev. B} \textbf{\bibinfo{volume}{20}},
  \bibinfo{pages}{3303} (\bibinfo{year}{1979}).

\bibitem[{\citenamefont{Laurent et~al.}(2006)\citenamefont{Laurent, Senes,
  Krebs, Kalevich, Urbaszek, Marie, Amand, and Voisin}}]{Laurent:2006a}
\bibinfo{author}{\bibfnamefont{S.}~\bibnamefont{Laurent}},
  \bibinfo{author}{\bibfnamefont{M.}~\bibnamefont{Senes}},
  \bibinfo{author}{\bibfnamefont{O.}~\bibnamefont{Krebs}},
  \bibinfo{author}{\bibfnamefont{V.~K.} \bibnamefont{Kalevich}},
  \bibinfo{author}{\bibfnamefont{B.}~\bibnamefont{Urbaszek}},
  \bibinfo{author}{\bibfnamefont{X.}~\bibnamefont{Marie}},
  \bibinfo{author}{\bibfnamefont{T.}~\bibnamefont{Amand}}, \bibnamefont{and}
  \bibinfo{author}{\bibfnamefont{P.}~\bibnamefont{Voisin}},
  \bibinfo{journal}{Phys. Rev. B} \textbf{\bibinfo{volume}{73}},
  \bibinfo{pages}{235302} (\bibinfo{year}{2006}).

\bibitem[{\citenamefont{Urbaszek et~al.}(2013)\citenamefont{Urbaszek, Marie,
  Amand, Krebs, Voisin, Maletinsky, H\"ogele, and Imamoglu}}]{Urbaszek:2013a}
\bibinfo{author}{\bibfnamefont{B.}~\bibnamefont{Urbaszek}},
  \bibinfo{author}{\bibfnamefont{X.}~\bibnamefont{Marie}},
  \bibinfo{author}{\bibfnamefont{T.}~\bibnamefont{Amand}},
  \bibinfo{author}{\bibfnamefont{O.}~\bibnamefont{Krebs}},
  \bibinfo{author}{\bibfnamefont{P.}~\bibnamefont{Voisin}},
  \bibinfo{author}{\bibfnamefont{P.}~\bibnamefont{Maletinsky}},
  \bibinfo{author}{\bibfnamefont{A.}~\bibnamefont{H\"ogele}}, \bibnamefont{and}
  \bibinfo{author}{\bibfnamefont{A.}~\bibnamefont{Imamoglu}},
  \bibinfo{journal}{Rev. Mod. Phys.} \textbf{\bibinfo{volume}{85}},
  \bibinfo{pages}{79} (\bibinfo{year}{2013}).

\bibitem[{\citenamefont{Schulhauser et~al.}(2002)\citenamefont{Schulhauser,
  Haft, Warburton, Karrai, Govorov, Kalameitsev, Chaplik, Schoenfeld, Garcia,
  and Petroff}}]{Schulhauser:2002a}
\bibinfo{author}{\bibfnamefont{C.}~\bibnamefont{Schulhauser}},
  \bibinfo{author}{\bibfnamefont{D.}~\bibnamefont{Haft}},
  \bibinfo{author}{\bibfnamefont{R.~J.} \bibnamefont{Warburton}},
  \bibinfo{author}{\bibfnamefont{K.}~\bibnamefont{Karrai}},
  \bibinfo{author}{\bibfnamefont{A.~O.} \bibnamefont{Govorov}},
  \bibinfo{author}{\bibfnamefont{A.~V.} \bibnamefont{Kalameitsev}},
  \bibinfo{author}{\bibfnamefont{A.}~\bibnamefont{Chaplik}},
  \bibinfo{author}{\bibfnamefont{W.}~\bibnamefont{Schoenfeld}},
  \bibinfo{author}{\bibfnamefont{J.~M.} \bibnamefont{Garcia}},
  \bibnamefont{and} \bibinfo{author}{\bibfnamefont{P.~M.}
  \bibnamefont{Petroff}}, \bibinfo{journal}{Phys. Rev. B}
  \textbf{\bibinfo{volume}{66}}, \bibinfo{pages}{193303}
  (\bibinfo{year}{2002}).

\bibitem[{sem()}]{semina}
\bibinfo{note}{M. A. Semina, to be published}.

\bibitem[{\citenamefont{H\"ogele et~al.}(2012)\citenamefont{H\"ogele, Kroner,
  Latta, Claassen, Carusotto, Bulutay, and Imamoglu}}]{Hogele:2012a}
\bibinfo{author}{\bibfnamefont{A.}~\bibnamefont{H\"ogele}},
  \bibinfo{author}{\bibfnamefont{M.}~\bibnamefont{Kroner}},
  \bibinfo{author}{\bibfnamefont{C.}~\bibnamefont{Latta}},
  \bibinfo{author}{\bibfnamefont{M.}~\bibnamefont{Claassen}},
  \bibinfo{author}{\bibfnamefont{I.}~\bibnamefont{Carusotto}},
  \bibinfo{author}{\bibfnamefont{C.}~\bibnamefont{Bulutay}}, \bibnamefont{and}
  \bibinfo{author}{\bibfnamefont{A.}~\bibnamefont{Imamoglu}},
  \bibinfo{journal}{Phys. Rev. Lett.} \textbf{\bibinfo{volume}{108}},
  \bibinfo{pages}{197403} (\bibinfo{year}{2012}).

\end{thebibliography}
\end{document}